\documentclass[prodmode,acmjetc]{acmsmall} 

\usepackage[ruled]{algorithm2e}
\renewcommand{\algorithmcfname}{ALGORITHM}
\SetAlFnt{\small}
\SetAlCapFnt{\small}
\SetAlCapNameFnt{\small}
\SetAlCapHSkip{0pt}
\IncMargin{-\parindent}

\begin{document}

\markboth{D Limbachiya et al.}{Natural Data Storage: A Review on Sending Information from Now to Then via Nature}

\title{Natural Data Storage: A Review on sending Information from now to then via Nature}
\author{Dixita Limbachiya
\affil{Dhirubhai Ambani Institute of Information and Communication Technology}
Manish K Gupta
\affil{Dhirubhai Ambani Institute of Information and Communication Technology}
}
\begin{abstract}
Digital data explosion drives a demand for robust and reliable data storage medium. Development of better digital storage device to accumulate Zetta bytes (1 ZB = $10^{21}$ bytes ) of data that will be generated in near future is a big challenge. Looking at limitations of present day digital storage devices, it will soon be a big challenge for data scientists to provide reliable. affordable and dense storage medium.  As an alternative, researcher used natural medium of storage like DNA, bacteria and protein as information storage systems. This article discuss DNA based information storage system in detail along with an overview about bacterial and protein data storage systems. 
\end{abstract}


\terms{DNA data storage, DNA codes, Bacterial based storage system, Biological Coding Theory, Error correcting codes}


\acmformat{Limbachiya Dixita, Gupta K Manish, 2015. Natural Data Storage: A Review on Sending Information from Now to Then}

\begin{bottomstuff}
Author's addresses: \\ 1.  Dixita Limbachiya \\ Laboratory of Natural Information Processing \\ Dhirubhai Ambani Institute of Information and Communication Technology\\ Gandhinagar, Gujarat, 382007 India\\
              Email: dlimbachiya@acm.org\\
              2. Manish K Gupta \\
              Dhirubhai Ambani Institute of Information and Communication Technology\\
			  Gandhinagar, Gujarat, 382007 India \\
              Email: mankg@computer.org
\end{bottomstuff}

\maketitle

\section{Introduction}
\label{intro}
\begin{figure}[ht]
	\centering
	\includegraphics[scale=0.38]{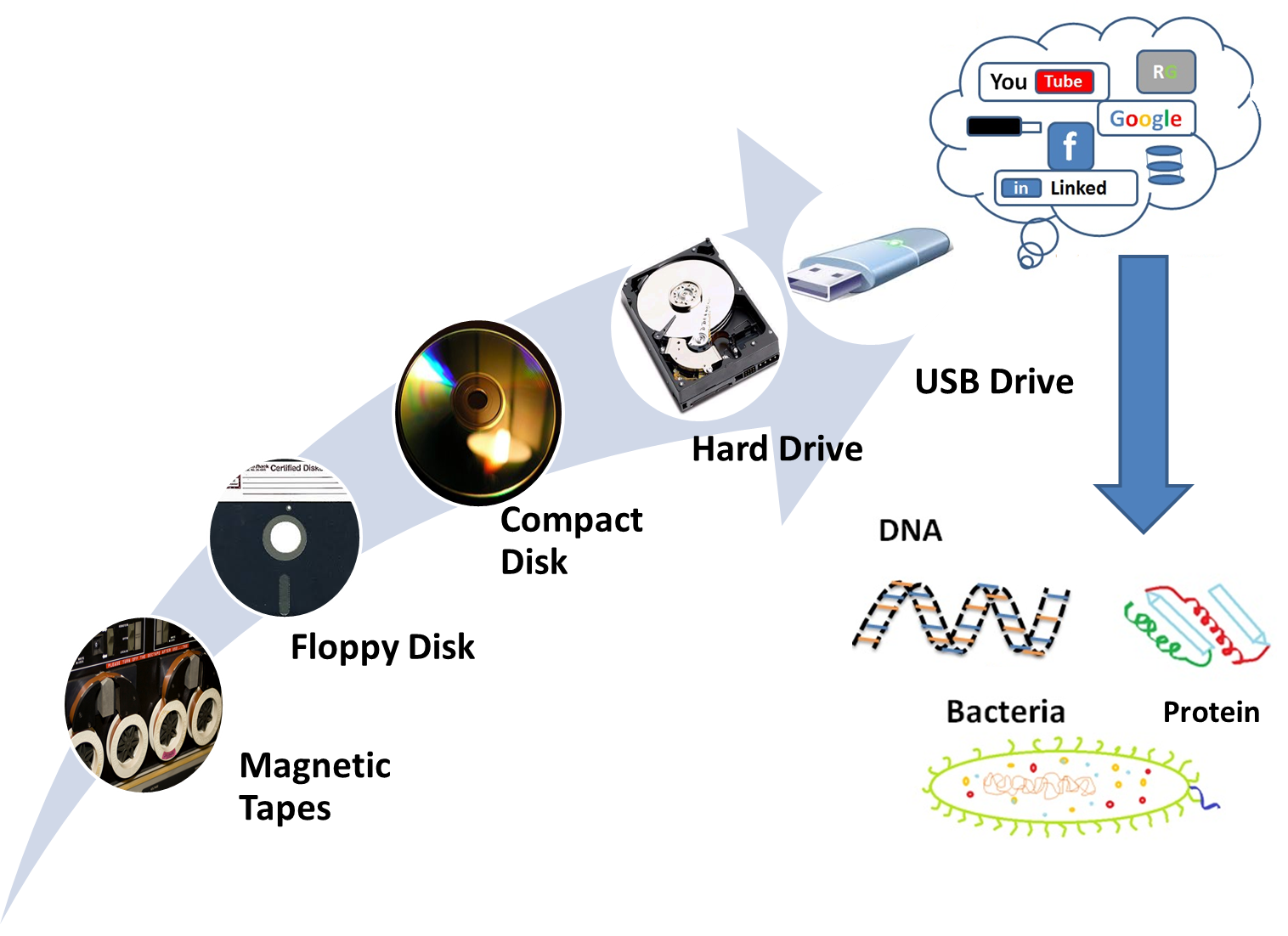}
\caption{Advancement in the field of data storage devices is shown here. New paradigm to store data on DNA, protein, bacteria is indicated.}
\label{nature}
\end{figure}

\begin{figure}[ht]
	\centering
	\includegraphics[scale=0.38]{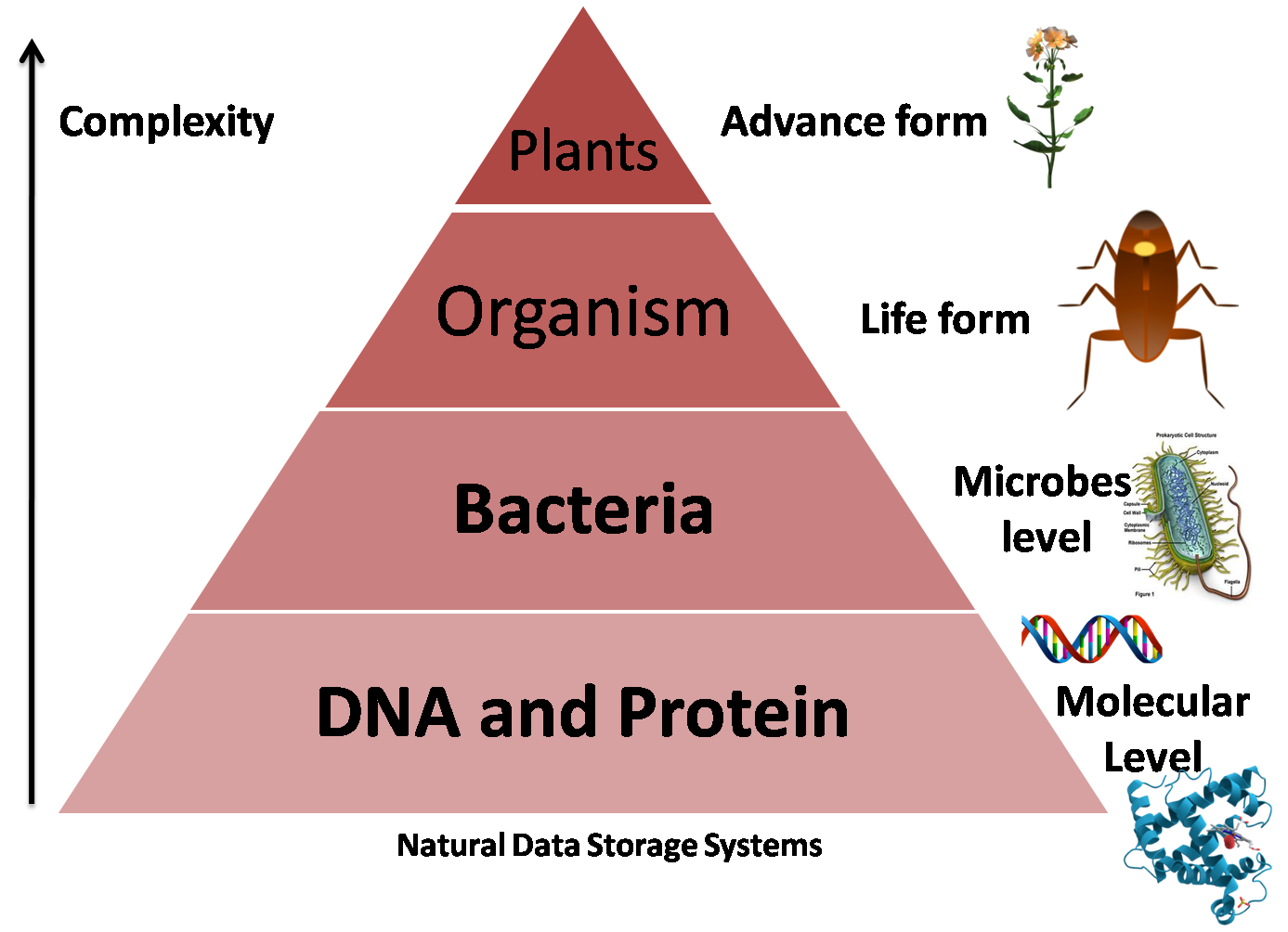}
\caption{Potential natural data storage medium and its complexity is indicated.}
\label{naturestore}
\end{figure}

With the extensive use of social networking and cloud computing, there is a paradigm shift in the volume of data produced. It is estimated that by $2020$, $35$ Zettabytes of digital information will be generated \cite{gantz2010digital}. This highlights a big concern of storing and maintaining the rapid growth of data that enforces the data storage experts to design a new architecture to store the data \cite{leong2012storage} \cite{du2008recent}. Steming from the early days storage medium like rocks, stones, paper, punch cards, magnetic tapes, CD, DVD, floppy disk,  etc. to the modern days distributed cloud data storage \cite{dimakis2010network}  \cite{dimakis2011survey} \cite{6463375}; there is a drastic advancement in the data storage devices (as depicted by Figure \ref{nature}). But these magnetic and optical discs are big, need to be maintained regularly and are prone to decay. Also they are not environment friendly as they consume vast amount of energy and release lots of heat. Scientist are trying to miniaturize the size of silicon chips up to many folds but this makes it more expensive. Alternatively, researchers instigate the use of living source from nature to preserve the data which give birth to biostorage. Biostorage is the field of storing and encrypting information in living cells or natural medium (see Fig \ref{naturestore}) \cite{baum1995building}. 
Comparison of natural storage medium and digital storage devices is appraised in \cite{mansuripur2002dna}. Many natural storage medium like DNA, protein, bacteria have been explored. This review paper showcase the evolution of DNA based storage systems along with its information encoding methods in details. Though there are evidences of reliable and scalable DNA based information systems, recent works by Church et al \cite{church2012next}, Goldman et al \cite{goldman2013towards} and \cite{grass2015robust} improved the efficiency of data encoding in DNA, which indicates the right time for the coding and information theorist communities to work on the challenges in natural data storage systems. Through this paper, one can witness the potential and challenges of DNA based information storage as well as other natural storage like bacterial and protein based storage systems. 

This review is structured as follows. Section $2$ introduce the DNA based information storage system. Section $3$ give brief description about bacterial data storage. Section $4$ describes the protein as hard drive. Section $5$ highlights the experimental evidence and challenges. This paper concludes with general remark.

\section{DNA as storage device}
\label{DNAstore}
\begin{figure}[ht]
	\centering
	\includegraphics[scale=0.38]{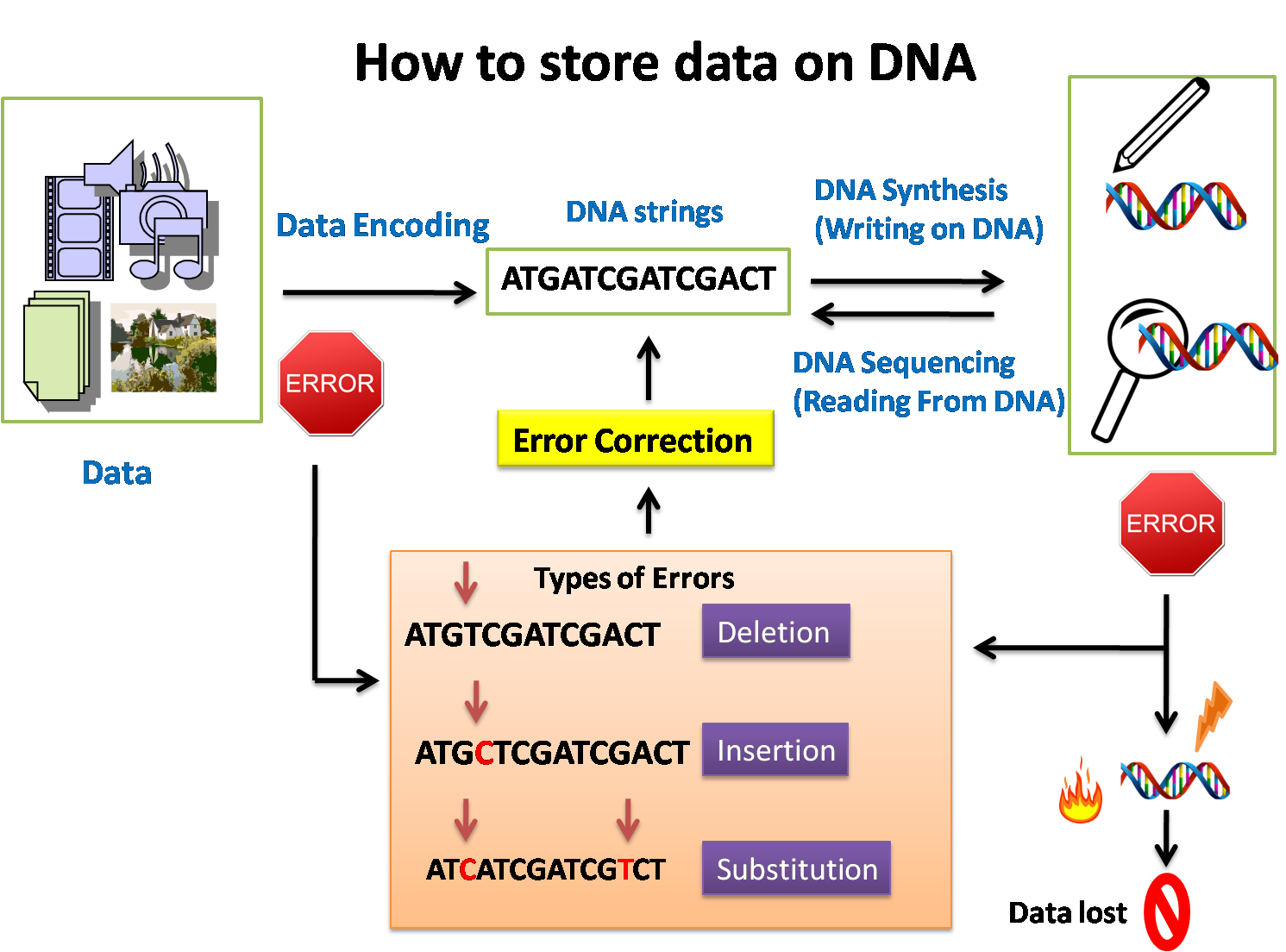}
\caption{Schematic representation of how to store data on DNA. Types of errors that may occur during DNA based data storage is depicted in the figure.}
\label{howtostore}
\end{figure}

\begin{figure}
	\centering
	\includegraphics[scale=0.50]{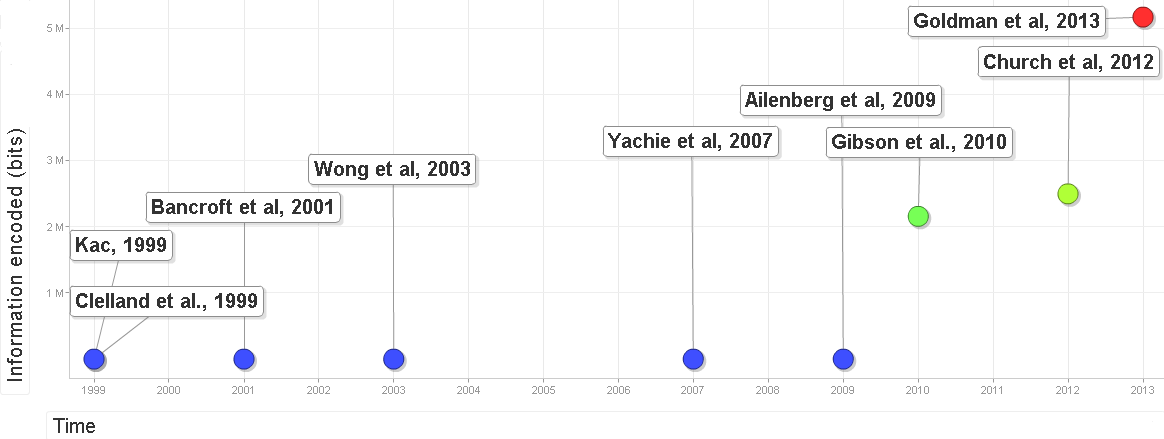}
\caption{Time-line for DNA based data storage systems.}
\label{timeline}
\end{figure}

DNA is natural information storage molecule, which stores our genetic information, is the favoured solution to the ample amount of data. DNA stores the genetic information using four bases A (Adenine), C (Cytosine), G (Guanine), and T (Thymine) analogous to digital storage device like CD which stores the information using lands and pits represented as 0's and 1's on the spiral tracks. 
The potential of DNA as a hard drive is well described in \cite{d2010comparative}. DNA is nature's hardware that has been used for computation which gave rise to DNA computation \cite{adleman1994molecular} \cite{PhysRevLett.80.417} \cite{amos1999theoretical} \cite{ezziane2006DNA} \cite{xu2007review} \cite{4696346}. DNA has a wonderful property of stability, long term storage, requires no electricity and needs no management beyond keeping it in a cold and dark place. DNA has self repair and error correction mechanism which has been witnessed by many researcher \cite{battail2006should} \cite{gupta2006quest}  \cite{milenkovic2006design} \cite{battail2007information} \cite{faria2012genome} \cite{Faria2014} which points to the capacity of DNA for the error correction. This opens the way to develop new encoding scheme which make sensible use of nucleotide base pairs. Initially DNA was used only to store text but with advancement in the field depicted in Figure \ref{timeline}, now DNA can be used to store any kind of data with $100$ percentage of data accuracy. Schematic diagram of how one can store data in DNA can be viewed in Fig \ref{howtostore}.

\subsection{Encoding data in DNA}\label{Encoding}
\begin{figure}[ht]
\centering
\includegraphics[scale=0.45]{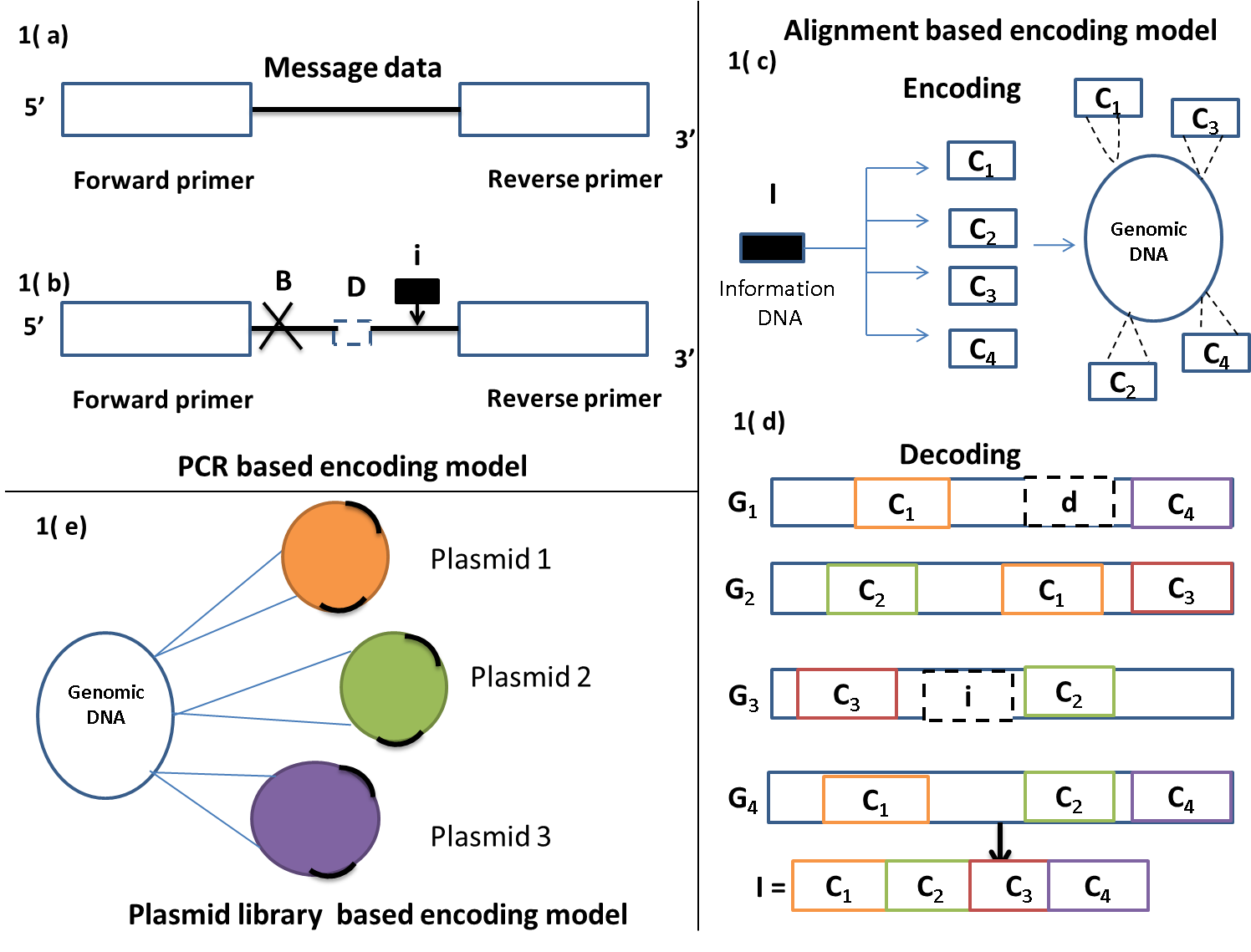}
\caption{PCR based, alignment based and primer library based encoding model is shown here. Fig 1(a) PCR based encoding model is template based method that consist of forward and reverse primer between which the information is inserted. Decoding is done by using this primers as template by PCR amplification . Fig 1(b) is the type of error that is base flip or break point indicates by B, deletion denoted by D and insertion of base denoted by i. Figure 1 (c) demonstrates the alignment based encoding model. Here I DNA sequence data that is converted to four different DNA chunks C1, C2, C3 and C4 using encoding principle.  DNA chunks are inserted at different loci in genome. For decoding (see fig. 1(d)), genome is sequenced and multiple alignment of different loci of genome is done for duplicated information inserted in multiple copies of genomic DNA. Figure 1(e) depicts encoding data in different plasmids in various location of bacterial genome. Data can be retrieved back by sequencing the index primers followed by sequencing of plasmid library.}
\label{encodingmodel}
\end{figure}

Data encoded in DNA can be used for encryption \cite{bancroft2006dna} or long term storage. Based on the purpose, DNA can be embedded in non-coding DNA (nc-DNA) or protein coding DNA (pc-DNA) or synthetic DNA. One can represent each base pair by using $2$ bits, which gives $4$ different possibilities that can be mapped to 16 combinations of DNA base pairs ( for instance $00$ $\rightarrow$ AT, $01$ $\rightarrow$ GC, $10$ $\rightarrow$ TA and $11$ $\rightarrow$ CG). A single byte (or 8 bits) can represent 4 DNA base pairs.  The entire diploid human genome can be represented in terms of bytes, as described: $6*10^9$ base pairs/diploid genome x $1$ byte$/4$ base pairs = $1.5*10^9$ bytes or $1.5 $ Gigabytes \cite{genomasize}. If we want to calculate data that can be stored in human body with consideration of human body consisting of $100$ trillion cells, we will have $150$ Zettabytes approximately ($150*10^{12}*10^9$ bytes) data stored in the DNA of any human. In the $20^{th}$ century, many researcher have translated English text, mathematical equations \cite{yachie2007alignment}, latin text \cite{portney2008length} and simple musical notations \cite{ailenberg2009improved} to DNA using different DNA coding principles \cite{wong2003organic} \cite{arita2004secret} \cite{skinner2007biocompatible} \cite{yamamoto2008large} \cite{heider2008dna}. Following are the main encoding approaches proposed for DNA based information storage systems as shown in Figure \ref{encodingmodel}.  

\subsubsection{Microvenus and Genesis project} 
Microvenus project was initiated by Joe Davis to convert image in DNA that allude the idea of storing a-biotic data in DNA. Microvenus \cite{davis1996microvenus}, a small organism comprises of short piece of synthetic DNA used to encode visual icon in bacteria \textit{E.coli}. Data encoding is done according to the molecular size of the bases. C is the smallest base assigned with 1, T$\rightarrow$2, A$\rightarrow$3 and G$\rightarrow$4. Instead of numeric values, each nucleotide was assigned with phase structure like C$\rightarrow$X, T$\rightarrow$XX, A$\rightarrow$XXX, G$\rightarrow$XXXX. The encoding was done by placing the nucleotide at each repeated position of bits 0s and 1s. Nucleotides were placed according to number of repeated bits of 0s and 1s. For instance, 1001011 = CTCCT, 10101= CCCCC. Mirovenus created was inserted into bacterial host cell by using plasmids. Encoding scheme used was not accurate, efficient and DNA developed to store data is not uniquely decodable. In subsequent year, other form of DNA based data encoding named Genesis \cite{Genesis}, an artwork of Eduardo Kac was introduced. He created artificial \textit{art} gene  that comprises of digital DNA by converting the lines from bible into Morse code. Morse code denoted by dot (.) and dash (-) was converted to nucleotides with the principle rule of converting dash (-) and dot (.) to T and C and replacing word space and letter space by A and G. This gene was then inserted into florescent \textit{E. Coli} bacteria. Both these laid foundation for storing data in DNA. But it lacks efficiency, accuracy and better encoding and decoding methods. 

\subsubsection{PCR based encoding models}
In Clelland encoding models \cite{clelland1999hiding}, microdots was used to cipher the data in human genomic DNA. Secret message was inserted between the PCR (Polymerase Chain reaction) forward and reverse primer sequences called template regions. The idea was to encode the characters by trivial assignment of DNA codon as encryption key to each characters and insert it in human genomic DNA. For total possible 64 codons, each English characters will be assigned codons and rest can be used to encode some of the symbols like dots and commas. This was carried out \textit{in vitro} by combining the message DNA with the genomic DNA in the solution over a 16-point microdot printed on filter paper. Decoding was based on template regions using PCR amplification. Recipient must be aware of encryption key and primer sequences. Data stored was secured but major limitation was scalability of data encoded in the limited size of microdots, only 136 bits data was encoded by using this approach. To make it more accurate, Bancroft \cite{bancroft2001long} proposed the concept of information DNA (iDNA) that comprises information and single poly primer key (PPK) along with forward and reverse primer and common 5-6 bases spacer to indicate the stored information. This concept resembles to the retrieval of information from an addressable storage device such as the random access memory in a computer where, PPK acts as data location identifiers. Data encoding was done by mapping ternary codes to only three bases A, C, and T but sequencing primers were designed with all four bases with the requirement that each fourth position be a G to prevent mispriming. Decoding can be done by sequencing PPK first to decode the forward and reverse primers and then based on specific sequencing primer one can retrieve the information. Total data encoded by using this approach was 561 bits. Drawbacks for PCR based methods are requirement of PCR, knowledge of primers and extensive experimental hurdles and practical issues. Moreover main drawback for PCR based methods are insertion of errors in template regions make the retrieval of encoded data impossible.

\subsubsection{Alignment based encoding models}
For the first time Yachie et al., \cite{yachie2007alignment} \cite{yachie2008stabilizing} introduced PCR independent alignment based data encryption using four bits per two bases encoding scheme. In this multiple sequence alignment based approach was used to encode the information into genomic DNA of \textit{B.subtilis}. Series of conversion of text to keyboard scan codes followed by conversion to hexadecimal code was followed to convert binary code to nucleotides with a designed nucleotide mapping four bits per two base. Multiple DNA oligonucleotides sequences carrying duplicated information was termed as cassettes and each cassettes were inserted redundantly into multiple loci of the \textit{Bacillus subtilis} genome. Data can be retrieved by multiple sequence alignment of the bit data sequences followed by genome sequencing without the need of template DNA or parity checks. Main drawback for the alignment based encoding was the size limit of the cassette oligonucleotides used to encode message. If it exceeds certain length, it will occur by chance in host genome. Also complete genome sequencing was needed to retrieve the data. All the models described, encoded text data in DNA, but in 2009 Ailenberg et al., \cite{ailenberg2009improved} proposed an improved Huffman coding with unique primer design, for the first time, they encoded text, images and musical characters in DNA. It employed modified base 3 Huffman code by dividing the keyboard characters into three groups of DNA codon and assigning DNA bases according the frequency of their occurrence. DNA sequences was synthesised and inserted in plasmid and decoding was done by sequencing of plasmid. Initially index plasmid which consist of information like title, authors, plasmid number and primers assignments is constructed. They used specific primers with unique prefix code for different types of file, for instance, text data was initiated by "tx" and music data was initiated by "mu". Data was inserted in plasmid with unique sequencing primer for information retrieval. They used nucleotides efficiently by encoding 4.9 bases per characters and encoded 1688 bits data.

\subsubsection{Church and Goldman encoding model}
Although aforementioned work were cornerstone for storing data to DNA they were successful on a small scale as it encoded small bits of data. The most rewarding work was done in recent times by Church, \textit{et al}. $2012$, at Harvard University . Using next generation synthesis and sequencing technology, Church came up with efficient one bit per base algorithm  of encoding information bits into fix length of DNA chunks (99 bases). Flanking primers at the beginning and end of information data was inserted to identify the specific DNA segment in which the particular data was encoded \cite{church2012next}. They encoded entire book (Regenesis: How Synthetic Biology Will Reinvent Nature and Ourselves ISBN-13:978-0465021758), including 53,426 words, 11 JPG images and one JavaScript program into 54,898 oligos each 159 nucleotide (nt) in length and consisting of a 96-bit data block (96 nt), a 19-bit address (19 nt) specifying the data block location and flanking 22 nt common sequences to facilitate amplification and sequencing. Initially the book to be encoded was converted into HTML format including all the images in it. The individual bits was converted to DNA sequence with conversion principle 1 bit per base encoding, A or C for 0 and T or G for 1. The bases were selected randomly avoiding homo polymer greater than 3 and constant GC content. The bits were indexed by 19 bits long bar-code sequence of consecutive number starting from $0000000000000000001$ which determines the location of encoded bits within the book. 
Each DNA segment was of length 12 without bar-code and the total number of oligonucleotides generated was $5.27$ MB. Specific primer sequence of $22$ nucleotide for the sequencing was designed and amplified using PCR. The sequence was read using an Illumina HiSeq next generation sequencer. In writing and reading DNA, $10$ bits error occurred from $5.27$ MB. It has only one drawback of lacking error correction scheme that was taken care by Goldman with $100$ percent of data retrieval.  

\begin{figure}
\centering
\includegraphics[scale=0.55]{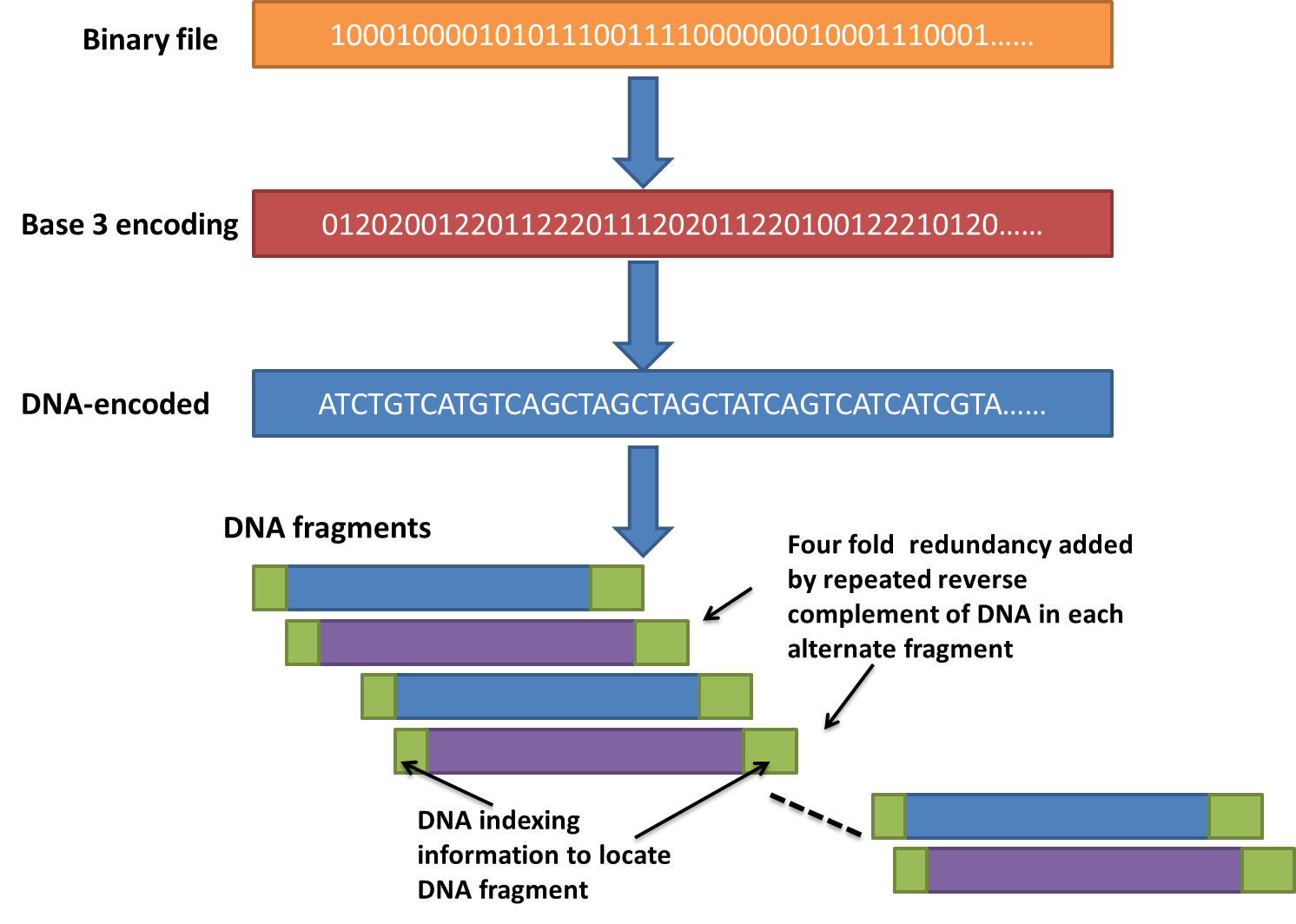}
\caption{Stepwise encoding of data into DNA using Goldman's approach is elucidated in detail. Binary data converted to base 3 Huffman code which then converted to DNA sequences. Each DNA sequences converted to fragments with each 75 base pairs overlapped in alternate fragment with reverse complement.}
\label{AdvancedDNASchemetic}
\end{figure}

In 2013 Goldman used one bit per base system introduced by Church and modifying it by employing the improved base 3 Huffman coding (trits 0, 1 and 2). In this original file in binary code (0, 1) is converted to a ternary code (0, 1, 2), which is in turn converted to the triplet DNA code. It involved four steps shown in Fig \ref{AdvancedDNASchemetic}. Binary digits holding the ASCII codes was converted to base-3 Huffman code that replaces each byte with five or six base-3 digits (trits). Each of trit was encoded with one of the three nucleotides different from the previous one used to avoid homo polymers that cause error in synthesis of DNA. DNA strand was divided into chunks each of length 117 base pair (bp). 75 bases for each DNA information chunks were overlapped with four fold redundancy to recover the data loss that occurred during synthesis and sequencing DNA. For the data security each redundant chunk was converted to reverse complement of the strand in every alternate chunks. Each DNA chunk was appended with data address blocks of 117 bases  to determine the location of segment in overall data. One parity check bit was added for the intra file location and error detection. Total 153,335 DNA strings were generated. 33 nucleotide base primer was added to facilitate synthesis process and amplification.  For details reader is referred to \cite{goldman2013towards}. As proof of concept, they used four different file types (739 kilobytes file size) and achieved 2.2 PB/g DNA storage density.



\subsection{Error correction in DNA based information systems}
There are different types of errors associated with DNA data storage systems which are physical errors and genetic errors. Physical errors occur during synthesis and sequencing of DNA and genetic errors are caused by mutations which occurs naturally during evolution and prolongation. Error can be insertion, deletion or substitution of single base in DNA sequences. Substitution of single base can be considered as bit flip errors. Other type of error can be deletion of bunch of DNA nucleotides categorized as burst error. Reading error rates ranges from 1-3 \% while writing error has error rates upto 15 \%. Error models proposed so far have focus on physical errors like substitution and deletion of oligonucleotides \cite{haughton2011repetition} \cite{KiahPM14} but no work has been done on insertion error model. 

There are three basic codes for storing data in DNA  \cite{arita2004writing} which are Huffman code, comma code and the alternating code. Although comma-free code and alternate codes are robust and has ability to correct against small-scale damage such as DNA point mutations, this cannot recover broken data block from the data-encoded DNA region. This breakpoints can be corrected by Huffman coding \cite{yachie2008stabilizing}. The DNA encoding by Huffman codes developed the Huffman coding method \cite{huffman1952method} is uniquely decodable. In this method, the probabilities of the symbol is considered (here the symbols are the English Alphabet). The least probabilities symbols are added to generate the next symbol and the process is repeated until we get the unique codes for all the symbols. For base 2 Huffman code, the least two  probabilities are added and reduced two generate compact code until all the symbols are coded with code $0$ and $1$. Likewise for base 3 and 4 base Huffman, least three and least four probabilities symbols respectively are reduced to compact code. The Huffman code described in \cite{smith2003some}, used the probabilities of the English alphabet. According to this the highest occurred letter \itshape e \normalfont has single base code and the least occurred \itshape z \normalfont has code length of 5 bases. For this base 2 Huffman code, the  two least frequencies are summed up to give compact code. The codes are assigned by varying the wobble position (third position of the codon) for the alphabet with similar probabilities. Average code length of the code is 2.2 which is shorter compare to other codes. The drawback of the method is that it do not include symbols and other characters. The improved Huffman code using all the English alphabets and special character was described using specific base assignment with uniquely designed primer sequences \cite{ailenberg2009improved}. The remarkable Huffman method was used by Goldman and his colleagues \cite{goldman2013towards} with error correction techniques.


In recent year, DNA storage channel was described by Han Mao Kiah et al; in which they represented reading and writing errors during DNA data storage as profile vectors and designed a family of error correcting codes for synthesis and sequencing errors \cite{KiahPM14}. They developed a codeword designed technique that resulted in the codes at sufficiently large distance that makes it best possible for error correction. DNA has essential property for long term archival of data compare to digital storage devices. To witness long term storage of DNA and improve the DNA stability, researchers have develop chemical based \cite{grass2015robust} method to encapsulate DNA into glass sphere and preserve it from environmental damage for long term archival. They used very prominent error correcting code Reed Solomon codes, which are used in digital storage devices like CD, DVD, to implement two layer of encoding one at DNA chunk level and other at synthesis and sequencing of DNA. This method can correct burst error. Recently re-writable DNA based data storage systems is proposed in \cite{rewritableDNAoligica}, in which they used unique addressing scheme by which data can be randomly accessed unlike previous techniques in which random access of data was not possible.

\begin{table*}[ht]
\tbl{Comparative study of the encoding schemes used to encode data in DNA with their limitations and constraints}{
\raggedleft
\tiny
    \setlength{\tabcolsep}{2pt}
\begin{tabular}{|l|l|l|l|l|l}
\hline
Source of storage&Encoding Scheme&Purpose&Limitation& Constraints\\ \hline \hline
Microdots & $4$ base encoding   & Data security  & No error correction used & Knowledge of PCR primer sequences \\
 &using PCR primers &and privacy&&and encryption is mandatory\\
 &and encryption key&&& \\ \hline
iDNA     & Poly primer sequence with & First implementation for actual & Security was not considered  & Information can be lost  \\
&designed encryption scheme &data stored in DNA using Microdot experiment& and not reliable&and designing of encryption key\\
&&&under adverse environmental condition&\\ \hline
Data encoded in  & 64 codon sequence mapped & Safe guard the data encoded in DNA inserted in vectors&Very few base pair information can be encoded &Preparation of Recombinant DNA and  \\
recombinant DNA & to ASCII characters uniquely&which can resist to adverse conditions & due to size limitation &designing vectors with proper sentinels\\
inserted in vector&& like radiations, extreme temperature&of genome size of &\\
&&&the vector (E.\textit{coli} and D.\textit{radiodurans}) used&\\ \hline
Plasmid Library& Huffman coding to &Storage of different file formats in DNA & Use 7 base encoding scheme & Designing of specific primer  \\
&English alphabets&&which may be ambiguous in decoding the data &sequence\\ \hline
Synthetic DNA& One bit per base- &Successfully stored and retrieved data from  & Lack in effective error correction & Ambiguity in the sequences \\
&0-A/C and 1- G/T & DNA&&\\ \hline
Synthetic DNA& Huffman code of base $3$  & Scaling the amount of data stored in DNA  & Time consuming for larger data & Knowledge of Synthesis\\
&to ASCII characters& with effective error correction&& and Sequencing of DNA\\ \hline
\end{tabular}}
\label{compareencode}
\end{table*}

\section{Shannon Information of DNA}
Shannon information \cite{shannon2001mathematical} for DNA is number of bits per DNA base. Theoretical limit of Shannon capacity of DNA is 455 Exabytes per one gram of DNA. It can be derived by considering 2 bits per nucleotide of single stranded DNA and average molecular weight of DNA $330.05$ g/mol/nucleotide. Shannon capacity of DNA can be obtained by calculating weight density per bit $(2.74 \times 10^{20}$ gram per bit), then calculating number of bytes in one gram of DNA $(1 \div 2.19 \times 10^{21}  = 4.55 \times 10^{20}$ bytes per gram of DNA). Different models are suggested based on errors for the Shannon capacity of DNA. One of the prominent idea about capacity of DNA was described by Vinhthuy Phan et al; \cite {phan2005capacity} in terms of hybridization model. He stated that it is very difficult to estimate the Shannon capacity for DNA of given length to store data. He considered hybridization can occur only if set of DNA are at some distance with parameter $\tau$ for the reaction stringency and gave the lower bound on DNA capacity to store the a-biotic data. Other models considered mutation in DNA sequences to estimate the Shannon capacity.  Capacity of DNA under the substitution  \cite {balado2010embedding}, insertion and deletion error \cite{balado2013capacity} was proposed by F. Balado. Considering encoding of data in non coding or coding region of genome, upper limit of Shannon capacity for amount of data stored in DNA under the error rate was specified. Upper bound for the DNA storage capacity without error is 2 bits per base. All the encoding methods used for embedding data in DNA have achieve Shannon information density ranging from 2 bits per base \cite{wong2003organic} to 0.213 bits per base \cite{heider2008dna} \cite{ailenberg2009improved} and 0.096 bits per base \cite{arita2004secret}. Goldman achieve 1.58 bits per base Shannon information for each DNA string. There are many low hanging fruits for designing the optimal capacity achieving codes for DNA storage with better code rate and length of DNA chunk. 

\section{Bacterial Hard drive}
\label{sec:3}
J.Cox \cite{cox2001long} suggested that suitable host to store data in DNA are \textit{Bacillus subtilis} and \textit{Saccharomyces cerevisiae} (bakers yeast). Yeast has higher density than bacteria but it is practically challenging. Bacteria has ability to survive in any condition like nuclear radiation, high temperature, deep under soil and water and in any hazardous condition. Potential bacterium that can be used for data storage are \textit{B. subtilis}, \textit{M.magneticum}, \textit{D. radiodurans }and \textit{Mycoplasma genitalium}. The idea of using bacteria as storage was pioneered by Masaru Tomita and Yachie \cite{yachie2007alignment} in which they stored famous Einstein relativity equation E MC Square in soil bacteria \textit{B.subtillius} to repeatedly store message DNA in multiple loci of genomic DNA of bacteria. They could encode 120 bits in 4.2 Mb genome of bacteria and decoded it back by multiple sequence alignment. 

Remarkable work was done by team of student at Chinese Hong Kong University using E.\textit{coli} as medium to store the data. It has storage capacity of $450 2,000$-gigabyte hard disks per gram of bacteria. This technology has many advantages over the magnetic data storage medium mainly information cannot be hacked and can defend against cyber attacks which points to higher data security than computer storage. An encoding system takes the original data, turns it into a quaternary number, and then encodes it as a DNA sequence by mapping {0,1,2,3} to {A,T,C,G} respectively with storage capacity of 1 Kb per cell. DNA sequences were compressed using deflate algorithm. This loss less data compression technique is important for two aspects, one is to increase the information storage capacity and other to avoid homo polymer ad repetitive regions in DNA. Information was broken down into fragments which consist of header sequence, message DNA and check sum. To retrieve the data, a novel biological information processing system was develop. Encryption is achieved through DNA sequence shuffling Rci recombination system by using site specific recombination by Recombinase (RCi) gene. They mapped the DNA using restriction enzyme so that data can be addressed just like filing system in magnetic storage. Live bacterial cells are used for data storage and they works like a transistor in the electronic devices which has on and off state. Memory device was designed that instruct the cells when to start the division and stop the division. This kind of devices will be useful in treatment of cancer and other diseases.

Storing the data in bacteria was a successful attempt but creating a rewritable storage was still a challenge that was solved by researchers at Stanford University by development of rewritable Recombinase addressable data (RAD) to store and rewrite digital information  \cite{bonnet2012rewritable}. With the help of enzymes one can modify DNA at specific site and can exchange DNA sequences at specific location. This can be done by enzyme recombinase which allows the strand exchange between site specific DNA sequences \cite{grindley2006mechanisms} which mimics the flipping behaviour of a bit by using recombinase-mediated DNA inversion \cite{ham2008design}. RAD module includes inversion of DNA by integrase and excisionase which depicts the bidirectional behaviour of the systems. It has two transcription input signals named set and reset. Set controls the expression of integrase that flips the DNA serving as data register. Other input reset drives the expression of integrase along with excisionase as co-factor that restores the direction of DNA element. This resulted into DNA registers which stores two states like finite automaton which can be flipped on basis of successive input signals. Here the states $0$ and $1$ are depicted by the green and red fluorescent protein, respectively. Depending upon the orientation of the specific DNA sequences the state of DNA is observed. Tuning and integrating the expression of recombinase, they developed a first reliable and rewritable DNA inversion-based data storage system.

Other perspective for bacterial data storage was perceived by researchers at Britain’s University of Leeds and Japan’s Tokyo University of Agriculture and Technology. They used bacteria \textit{Magnetospirilllum magneticum} to organically grow tiny magnets which can store bits of data \cite{SMLL:SMLL201102446}. Magnetic storage devices are built by cutting down the large magnets into tiny magnets and embedding it on to the storage tapes like hard disk and DVD. Instead of cutting magnets, scientist thought of creating tiny magnets out of some natural source that can be used to store the data. This bacteria has capacity to ingest the iron and produce the crystals of mineral magnetite. Researchers studies this mechanism of the bacteria and tried to mimic the same outside the body. They arranged magnetic particle specific iron-binding protein Mms6  \cite{galloway2012nanomagnetic} in chess board pattern and dipped in the iron solution \cite{tanaka2011mms6}. This experiment resulted in growth of tiny magnets that may be used as potential material to built the storage devices \cite{SMLL:SMLL201101627}. Each nano cube can store a bit of information with size $20000$ nano meters wide much larger than magnetic storage device with $10$ nm magnetic pits. The researchers are now working on miniaturizing the size of nano magnets and using alternative magnetic material to develop single array nano magnet that can store one bit of information. In-spite of these success stories for bacterial data storage, there are many problems to be focused. Designing of error models for bacterial data storage, development of error correction codes, modeling of Shannon capacity for bacterial storage are main difficulties.

\section{Protein hard drive} 
\label{sec:2}
Protein plays central role in the functioning of the body and stores the behaviour of human in form of folded chain of amino acids. The communication takes place between various organs at particular instance of the time in response to specific protein. In order to use protein as the storage medium, identification of proteins which can replicate the binary storage technique was inevitable. The study of photo-switching protein has revealed many applications in nanoscience technology one of which is data storage \cite{Sauer05072005}. The foundations for the same was laid by Hirschberg and colleague by proposing first photochemical memory model based on the color transformation of molecule called spiropyrans which can flip to other form on absorption of single photon \cite{doi:10.1021/ja01591a075}. Furthermore, using photo-induced fluorescence proteins that can switch between two states for data storage is described in \cite{tsien1998green}. Switching mechanism can be between the state of two different colors like red and green, other can be dark or light state. Researchers in the area found solution to it by studying family of proteins called photochromic proteins like Photo convertible Florescence proteins (PCFPs) and Reversible switching florescence proteins (RSFPs) which are light driven switchable florescent proteins \cite{Adam2010289}. Not only reading and writing the data but they could erase the data and rewrite it again on the proteins. So this was first remarkable attempt of creating rewritable natural data storage. PCFPs protein called Kaede \cite{ando2002optical} and RSFPs protein known as Dronpa \cite{ando2004regulated} was used to write and rewrite the data. Cis/trans isomerization of chromophore \cite{Lukyanov25082000} along with photon induces protonation \cite{Adam25112008} of chromophore is responsible for photo-switching in two different states. The information can be stored in the area designated green and red colors which are similar to 0 and 1. The state was determined by using EosFP, a fluorescent marker protein which is UV-inducible green-to-red fluorescence. Different material as described in \cite{Adam2010289} were used for surface coating of all proteins. To write data on the protein surface, an inverted laser-scanning microscope with particular specification was used. The reading, writing and erasing of data was done by using the laser beam at different intensity levels.

The idea of using single crystals of PCFPs/RSFPs protein as 3D storage medium had been implemented where protein molecule in the crystal would represent a data bit \cite{hell2007method}. In this instead of binary encoding, four color florescence switching proteins was used with mutant of RSFPS which can built the 4-base data storage system by using two photon excitation (TPE) technique \cite{mandzhikov1973nonlinear}. Using the mutant of PCFP EosFP, IrisFP \cite{Adam25112008} the 4 based data storage was implemented as it had combine property of both the proteins of irreversible switching from green to red and other reversible from dark to bright state. This technique is more efficient as it focus on the precise location for the storage. 

One other idea was implemented by group of Venkatesan Renugopalakrishnan at Harvard University using the Bacteriorhodopsin (bR) protein for data storage. The concept of using bR was spearheaded by Jack tallent \cite{tallent1996effective}. The researchers used Bacteriorhodopsin (bR) light-activated protein found in the membrane of a salt marsh microbe \textit{Halobacterium salinarum} (which use it for their photo synthesis) to coat storage device DVD which may increase the data storage capacity up to $50$ TB \cite{4143373} \cite{oesterhelt1991bacteriorhodopsin} \cite{renugopalakrishnan2006protein}. It convert light energy into chemical energy. When lights comes in contact it produces some molecules of different state and remain in the state for few minutes or days \cite{birge1995protein}. Unlike today's storage device the bR molecule of $~2$ nm in diameter has demonstrated a long-term stability with a shelf life of at least 10 years at room temperature and is believed to be stable at temperatures up to $140$ degree Celsius. The team modified bR protein to enhance its thermal and photochemical properties in a way that it can remain in this intermediate state for few years at high temperature \cite{renugopalakrishnan2003retroengineering}. In this the binary encoding is done by the concept that protein in brighter state is considered as $0$ and in dark state can be considered as $1$. They worked on how to use charge transporting proteins such as Bacteriorhodopsin in the building of data storage and transmission devices for applications in computer technology. When laser of one color incidents on the protein, it get arrange into one shape designated as zero in binary system and when laser of another wave length stimulates the protein to take another shape represented as one. Once the laser system is switched off, data can be stored for several year. To read the data from stored protein, a low power laser beam is delivered on this protein slowly so that the protein confirmation is not disturbed but only the light is absorbed by the pattern in the protein which can be detected by the machine and can generate a string of 0s and 1s. The property of bR to shifts between intermediated states made it potential for rewritable data storage. 

Using peptide as storage device was implemented by amalgamation of the nm scaled bio-organic nano dots into bio-electronic devices \cite{amdursky2013bioorganic}. This bio-organic nano-dots called peptide nano dots (PNDs) of $2$ nm size, were self-assembled from the Diphenylalanine (FF) peptides  \cite{jeon2013molecular} and can be embedded into metal-oxide-semiconductor devices as charge storage nano-units in non-volatile memory. FF gets self assemble into nano tubes and the structure is stabilized by the non-covalent interactions \cite{santhanamoorthi2011diphenylalanine}. Size limitation (micrometer) of the FF tubes jeopardized it's use for the bio-electronic devices. But researcher observed that in anhydrous condition, FF tubes get dissemble into stable building blocks of PNDs, paving the way for using it in bio-electronic devices. PNDs were successfully used as charged storage elements for the non volatile memory devices (NVM) by replacing the ONO(oxide-nitride-oxide) dielectric in the NVM. Many other nano-dots like Au and Pt and organic dyes were earlier employed for the same, but the beauty of PNDS to be nano-crystalline, uniform nm size, low temperature deposition makes them superior. Two crucial steps were followed to use FF PNDs in NVMS. First the property of each nano-dots was deciphered using electron microscopy. Using their electron diffraction pattern, nano crystalline structure was confirmed. Following it mono layer of PNDs which serves as memory stack was formed which retains the charge. For the detail procedure, reader is advised to refer \cite{amdursky2013bioorganic}. Protein bases information storage systems are at infancy stage and opens many challenges like reading and writing the data, data rate, speed of data access and deciphering the mutants of the photo-chromic proteins for intense research.

\section{Experimental Evidences and Challenges}

Research done so far,undoubtedly,acquaint about the potential of natural data storage devices but still to bring it into commercial applications, many issues like cost for the storage and experimental challenges and human expertise are to be perceived. Today storage device can read data at $100$ MB per sec, which is a much higher than the data access rate of nature hard drive. Despite of the fact that DNA is scalable, stable and robust storage device, synthesis and sequencing process involved are time consuming and require the expertise which make DNA storage an unreachable to commoner. As development in the field of next generation sequencing techniques is accelerating at higher rate compare to the digital storage medium. It can be estimated that in near future this technology will become cheaper \cite{schatz2010cloud}. Decreasing the cost of the synthesis and adapting parallel automation \cite{cleary2004production} and simplified purification techniques are the aspects one has to focus. Improved error correction schemes with high storage capacity are the theoretical challenges in the field. Other limitation is size of the DNA fragments that is used to store this data. Current synthesis and sequencing techniques are limited to process certain small size of DNA sequences. So the advancement in the sequencing and synthesis techniques can aid to make the DNA storage more feasible in coming era \cite{fuller2009challenges} \cite{IJCH:IJCH201300032}. Taking the first step towards the construction of molecular data storage system, researchers have made a paradigm shift in DNA reading and writing techniques by proposing the technique to built a DNA storage device that has reading and writing chambers \cite{khulbe2005dna}. DNA readout experiments conducted in miniaturized chambers which can be alternative to existing technology for DNA synthesis and sequencing have been explored. They adapted the methods of DNA processing used by molecular biologist to built the data storage chip. String of macromolecule (here DNA) containing the bytes of information is created and to secure the information they are translocated to safety zones called parking spots. To read the data, they may be transferred physically to decoding stations and data can be by controlled electric-field gradients, electronic micro motor etc. \cite{mansuripur2005information}. Obstacles in this methods are data access and data rate that is very low (12 kbit/s) that is to be improved. For more details and system architecture reader is suggested to refer \cite{khulbe2005dna}. 
 The other important challenge is the ease retrieval and random access. It need efficient random access and improved rewritable methods. Scaling the natural storage capacity is one of the important area of research to make molecular storage as commercial application. For the data stored in bacteria, the bacteria cultures and incubation required a lot of human expertise to avoid a chance of contamination. This rise an issue of data security. So this methodology has to deal with data security and knowledge base to handle the population of bacteria used as the medium to encode data. Moreover there are many low hanging fruits in the area of encoding and decoding algorithms to store data in  bacteria. With the bliss of synthetic biology, scientist at Craig Venter Institute  have synthesised artificial bacterial cell with synthetic chromosome and watermarked data into living cell of bacterium \textit{Mycoplasma mycoides} which has capacity of self replication \cite{Gibson02072010}. Novel innovation like this motivates the dream of bacterial data storage. As far as protein is concern, there is lot be explored. There is very few evidences which depicts the potential of protein as storage medium but the above mentioned work is headway for new milestones. Breakthroughs in programmable protein synthesis which may replicates the nature of computer hard drive is at far vision in the domain. Efficient algorithm which can map data to amino acids sequence is one most challenging part to make this possible. Below is the executive summary of the natural data storage and their challenges.

\begin{table*}[ht]
\tbl{Executive Summary of natural data storage devices and their evidences}{
\raggedleft
\tiny
    \setlength{\tabcolsep}{2pt}
\begin{tabular}{|l|l|l|l|l|}
\hline
Source &Encoding and decoding algorithm&Storage capacity achieved&Properties& Experimental validation\\ \hline \hline
DNA   & Yes, effective Huffman base  & 490 EB (exabytes) per gram & Rewritable, scalable, stable  & Yes, data was stored \\
	&3 encoding scheme&&under extreme conditions, dense& and retrieved with 100 percent accuracy\\ \hline
Bacteria & Yes, but preliminary encoding   &$450 2,000$-gigabyte hard  & Rewritable, secured, dense, & Yes, (as mentioned in \\
&data to codes and mapped to bacterial genome& disks per gram of bacteria& high duplication rate&section Bacterial hard drive)\\ \hline
Protein    & Yes effective encoding data to colors, proteins & Yet to be uncover& Long term storage, secured, stable&Yes, as described above  \\
&with two state systems (dark-bright; &&&in section Protein hard drive\\
& green-red; or both for 4-base data encoding)&&&\\
&are used to mimic the binary storage system in computer&&&\\ \hline
\end{tabular}}
\label{comparedata}
\end{table*}

\section{Conclusion}
With this explosion in the amount of data, natural storage seems to be the solution to preserve the data as archival for longer period. Considering the challenges for the natural data storage, it will not immediately replace the computer storage drives. Nevertheless, with the advancement in synthetic biology technologies, the day is near where this dream will come true.  The main focus of this area can be on data security, improved encoding and decoding approaches, making technology cost effective and far-sighted developing the protein and molecules of bacteria to generate the components of the computer by using bottom up approach. The quest for improving the existing storage devices craving for the energy has forced the researchers to turn their attention to replace it with eco-friendly storage devices in coming decade. Though there are many rooms for the development of robust natural storage device, one can imagine a near future where the technology will allow the computers around the internet to exchange the information on its own, self replicate the information and even mutate or improve the content and correct error on its own. 


\bibliographystyle{ACM-Reference-Format-Journals}
\bibliography{dnastoreref}


\end{document}